\documentclass[preprint,12pt]{elsarticle}

\usepackage{amssymb}
\usepackage{amsmath}
\usepackage{lineno}
\usepackage{siunitx}
\DeclareSIUnit\barn{b}

\journal{Nuclear Physics B}

\begin{document}

\begin{frontmatter}



\title{{LF-MightyPix}: A second {HV-MAPS} prototype for the {LHCb} {Mighty-Tracker}} 

\author[KIT]{Toko Hirono}
\author[HD]{Sebastian Bachmann}
\author[HD]{Lucas Dittmann}
\author[KIT]{Richard Leys}
\author[KIT]{Nicolas Striebig}
\author[HD]{Celina Welschoff}
\author[KIT]{Hui Zhang}
\author[KIT]{and Ivan Peric}

\affiliation[KIT]{organization={Institute for Data Processing and Electronics, Karlsruhe Institute of Technology},
             addressline={Hermann-von-Helmholtz-Platz 1},
             city={Eggenstein-Leopoldshafen},
             postcode={76344},
             country={Germany}}
\affiliation[HD]{organization={Physikalisches Institute, University of Heidelberg},
            addressline={Im Neuenheimer Feld 226},
            city={Heideberg},
            postcode={69120},
            country={Germany}}

\begin{abstract}
For the future high‑luminosity operation of the LHCb experiment, the downstream tracker will be upgraded to the Mighty-Tracker.
A key part of this upgrade is the introduction of silicon pixel detectors, MightyPix, in the central region of the tracker.
We have developed MightyPix prototype chips using High‑Voltage Monolithic Active Pixel Sensors fabricated in a commercially available CMOS process on high‑resistivity wafers. 
The second prototype chip, LF-MightyPix, is fabricated in the LFoundry 150~nm CMOS process.
LF-MightyPix has a chip size of \mbox{\SI{3.5}{\milli\meter} \(\times\) \SI{4.0}{\milli\meter}} and a pixel size of \mbox{\SI{100}{\micro\meter} \(\times\) \SI{100}{\micro\meter}}.
For each pixel hit, both the time of arrival and the time over threshold are recorded to ensure correct bunch-crossing identification at \SI{40}{\mega\hertz}.
The results presented in this paper confirm compatibility with the MightyPix requirements.
\end{abstract}



\end{frontmatter}


\section{Introduction}
\label{sec:intr}

The Large Hadron Collider beauty (LHCb) experiment \cite{The-LHCb-Collaboration-2008} at CERN after its Upgrade II, scheduled for 2034--2035, is expected to achieve an instantaneous luminosity of up to \SI{e34}{\per\centi\meter\squared\per\second} and to collect a total integrated luminosity of \SI{300}{\femto\barn^{-1}} \cite{LHCb-TDR-023}. 
To operate under this high-luminosity condition, the downstream tracker currently based on scintillating fibers will be upgraded to the Mighty-Tracker, whose central region will be based on silicon pixel detectors named MightyPix \cite{LHCbScoping}.
MightyPix is based on High Voltage Monolithic Active Pixel Sensors (HV-MAPS) \cite{HVMAPS-Peric} since the monolithic design simplifies construction compared to hybrid sensors, which is advantageous for large-scale detector systems such as the Mighty-Tracker.

Due to the complexity of MAPS design, both the sensor and the readout electronics require careful optimization.
The full development of MightyPix is expected to take more than ten years from initial planning to final installation.
The discontinuation of the CMOS process during the development could delay the project schedule.
Therefore, evaluating alternative CMOS processes as potential backup options is important to ensure long-term feasibility.

In this paper, we describe the prototype LF-MightyPix, which was ported to an alternative CMOS process technology.
The suitability of this technology is evaluated in terms of sensor depletion and timing performance at the target power consumption.

\section{MightyPix}
\label{sec:mp}

\subsection{Requirements and prototype development}
\label{sec:proto}

Requirements for the MightyPix, derived from the physics goals of the LHCb Upgrade~II, are described in earlier studies \cite{LHCb-TDR-023,LHCbScoping}. 
These requirements are summarized in Table~\ref{tab:req}.
The MightyPix development includes three prototype chips prior to Mighty-Tracker production: MightyPix1, MightyPix2, and MightyPix3.
MightyPix1 is the first prototype dedicated to the Mighty-Tracker and implements the basic LHCb control and data-acquisition interfaces. The chip has been fabricated and tested \cite{Scherl2025}.
MightyPix2, with a chip size of \SI{20.1}{\milli\meter} \(\times\) \SI{18.7}{\milli\meter}, integrates nearly all functions foreseen for the final design \cite{Striebig2026}.
The chip has been submitted for fabrication.
MightyPix3 will be the final full-reticle pre‑production prototype and will include improvements based on the results of the previous prototypes.

\begin{figure}[h]
\centering 
\includegraphics[width=0.9\columnwidth, trim=20 0 10 0, clip]{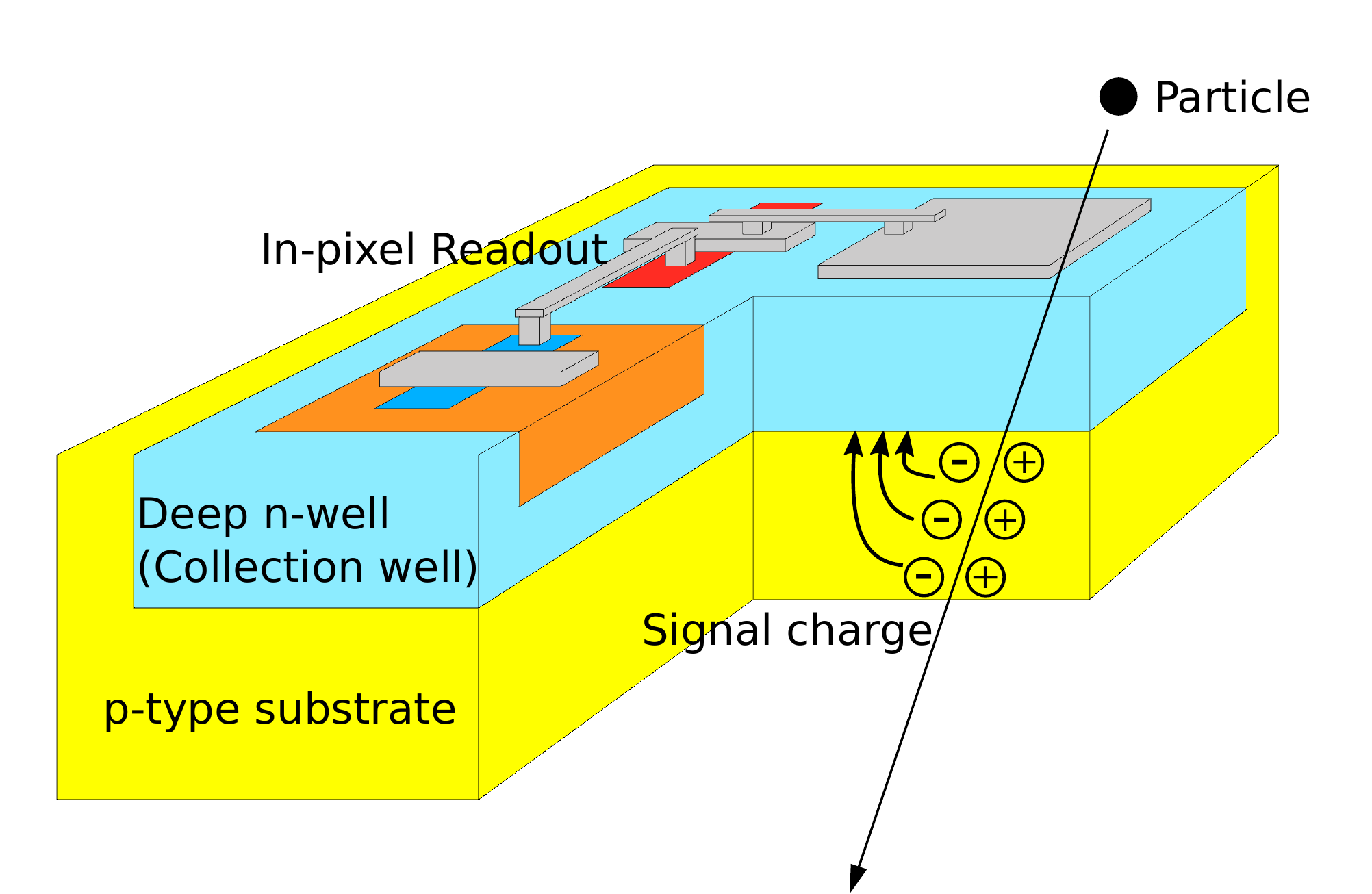}
\caption{Simplified diagram of HV-MAPS structure.}
\label{fig:hvmaps}
\end{figure}

The structure of HV-MAPS is shown in Fig.~\ref{fig:hvmaps}.
A charge-collection well is implemented as a deep n-well, the deepest n-well available in the process.
A p–n junction forms between the collection well and the p-type substrate, where a reverse bias voltage will be applied.
Multi-well structures enable CMOS readout electronics to be inside the collection well and isolated from the substrate.
Thus, the HV-MAPS design allows a high bias voltage to be applied to the substrate.

To meet the MightyPix requirements, CMOS processes are selected based on the following criteria: 
1) support for multi-well structures, 2) availability of high-resistivity wafers, 3) sufficient production capacity for the Mighty-Tracker, and 4) demonstrated radiation hardness in previous projects.
Criterion 2) is necessary to achieve the required radiation tolerance. 
A sufficiently large depletion region, formed by the high substrate resistivity and a high bias voltage, enables fast charge collection by drift before radiation-induced charge trapping occurs \cite{Wermes2020}.
The TSI/AMS 180~nm CMOS technology \cite{tsi,ams} used for MightyPix1 and MightyPix2 satisfies all these criteria, including a substrate resistivity of \SI{> 280}{\ohm\centi\meter} and demonstrated radiation hardness \cite{Benoit2018}.

\begin{table}[h]
    \centering
    \begin{tabular}{p{0.4\columnwidth} p{0.45\columnwidth}}
    \hline
    Parameter                     & Value \\                                                                  
    \hline
    Area to be covered            & \(\sim\)\SI{18}{\meter\squared} \\                                          
    Pixel size                    & \(\leq\) \SI{100}{\micro\meter} \(\times \leq\) \SI{300}{\micro\meter} \\   
    Chip thickness                & \(<\) \SI{200}{\micro\meter} \\                                             
    Maximum hit rate              & \SI{34}{\mega hit\per\second\per\centi\meter\squared} \\                    
    Overall detection efficiency  & \(>\) \SI{96}{\percent} including inactive areas (e.g.\ chip gaps) \\       
    In-time detection efficiency  & \(>\) \SI{99}{\percent} within \SI{25}{\nano\second} for the active area \\ 
    NIEL (End of Life)            & \SI{3e14}{(1MeV) n_{eq}\per\centi\meter\squared} \\                         
    TID (End of Life)             & \SI{40}{\mega rad} \\                                                       
    Power consumption             & \(<\) \SI{150}{\milli\watt\per\centi\meter\squared} \\                      
    Data transmission rate        & 1.28~Gbps per port \\                                                       
    \hline
    \end{tabular}
    \caption{Requirements of the MightyPix}.
    \label{tab:req}
\end{table}

In parallel, LF-MightyPix has been developed as a prototype using the LFoundry 150~nm CMOS technology \cite{lfoundry}.
It also meets the previous criteria, including wafers with resistivity above \SI{2}{\kilo\ohm\centi\meter} and proven radiation hardness \cite{Hirono2020}.
LF-MightyPix has been designed taking into account the differences between the two CMOS processes, such as the wafer resistivity (\SI{> 2}{\kilo\ohm\centi\meter} and \SI{> 280}{\ohm\centi\meter}), feature size (150~nm and 180~nm), and the well structure details.

\subsection{LF-MightyPix}
\label{sec:lfmp}

\begin{figure}[h]
\centering 
\includegraphics[width=0.9\columnwidth, trim=0 100 300 100, clip]{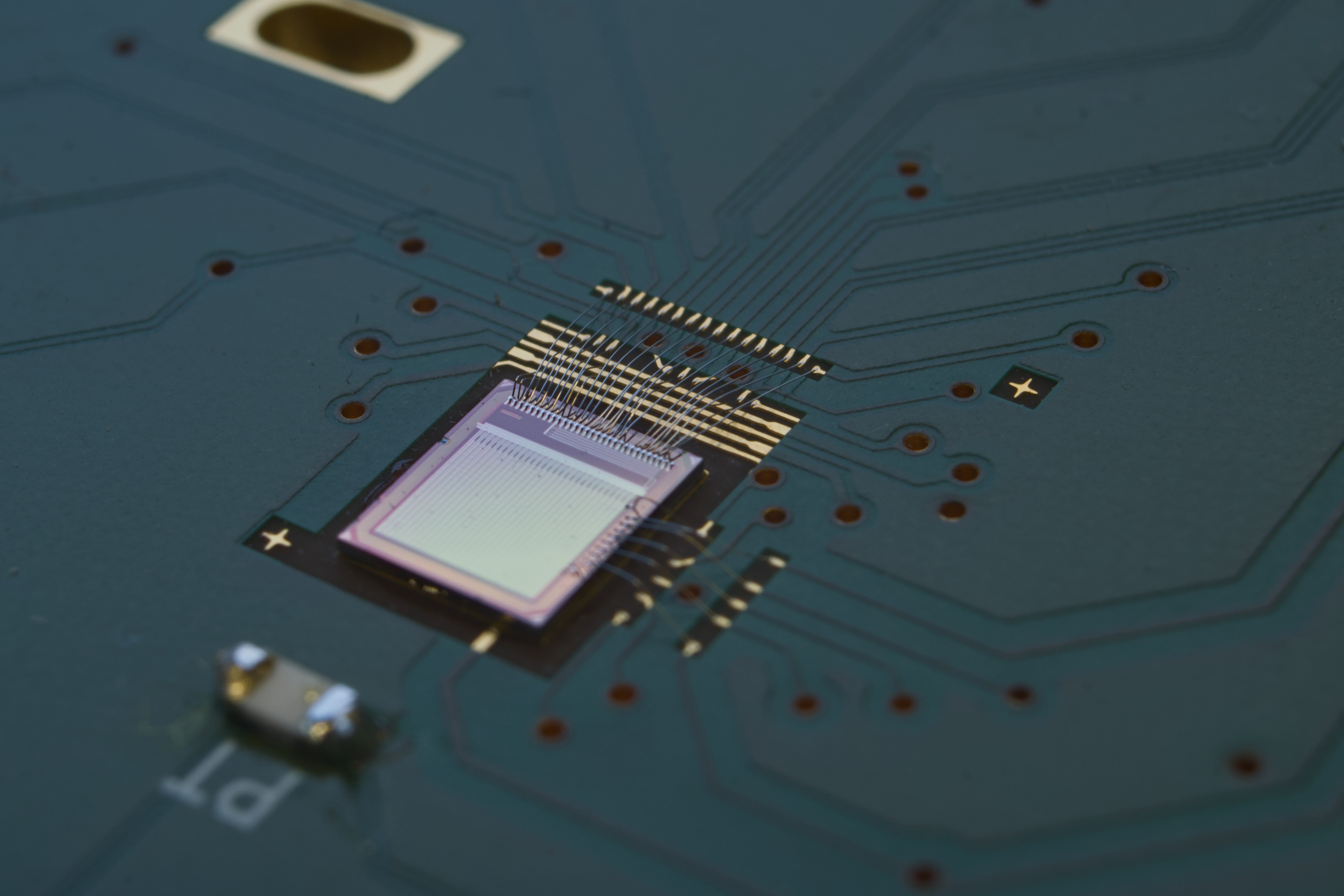}
\caption{LF-MightyPix on a PCB for the GECCO system.}
\label{fig:lfchip}
\end{figure}

\begin{figure*}[h] 
\centering
\includegraphics[width=\textwidth, trim=10 50 10 10, clip]{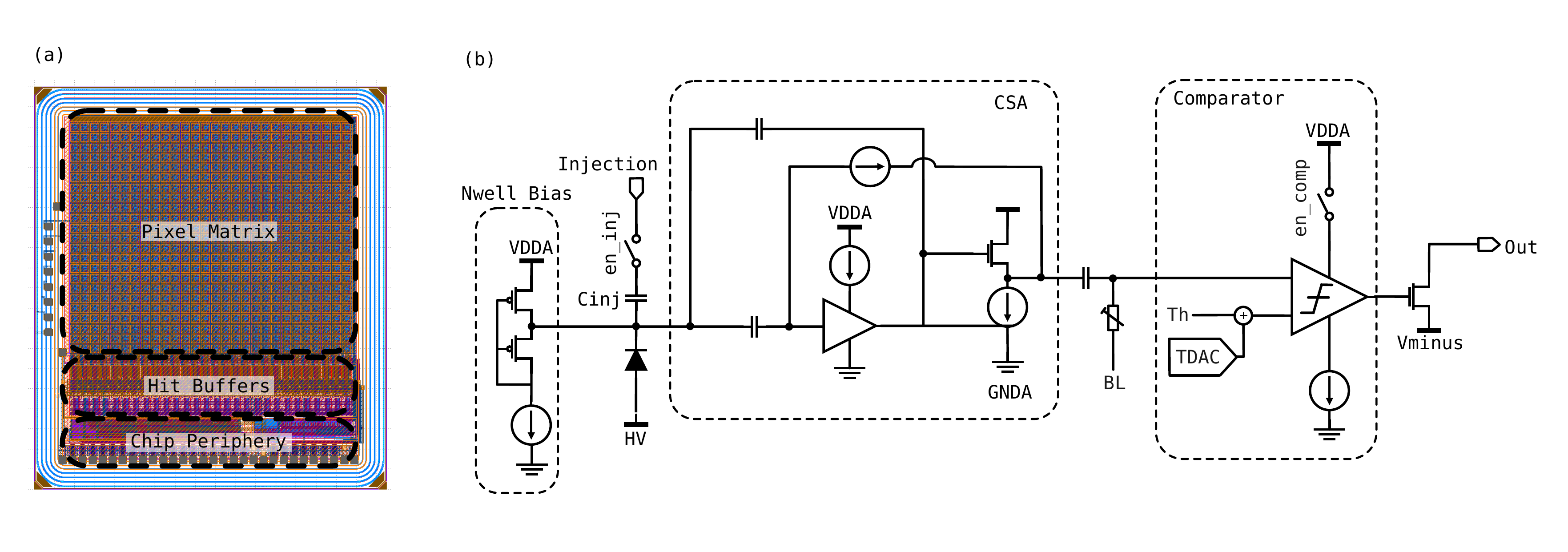}
\caption{Chip layout (a) and in-pixel readout circuit (b) of LF-MightyPix.}
\label{fig:pix_ana}
\end{figure*}

LF-MightyPix has been fabricated (Fig. \ref{fig:lfchip}) as a small prototype chip with dimensions of \mbox{\SI{3.5}{\milli\meter} \(\times\) \SI{4.0}{\milli\meter}} and a thickness of \SI{280}{\micro\meter}.
It consists of three main components (Fig. \ref{fig:pix_ana}a).
The largest component is the pixel matrix, which has 28 columns and 23 rows.
The pixel size is \mbox{\SI{100}{\micro\meter} \(\times\) \SI{100}{\micro\meter}}.

The in‑pixel readout electronics include n-well biasing circuits, a charge\-sensitive amplifier (CSA), and a comparator (Fig.~\ref{fig:pix_ana}b) \cite{Peric2021}.
The n-well biasing circuits apply a voltage of approximately \SI{1.8}{V} to the collection well.
A capacitor connected to the collection well (\(\mathrm{C_{inj}}\)) enables test-pulse injections.
The CSA amplifies the collected charge and converts it into a voltage signal.
A high‑pass filter between the CSA and the comparator shapes the signal before discrimination.
Each pixel includes a 4‑bit trim DAC (TDAC) and a 1‑bit switch, and their configuration bits are stored in a 5-bit RAM.
The TDAC adjusts the comparator threshold, and the switch enables or disables the comparator output.

The comparator output is sent to the hit buffers, which are located between the pixel matrix and the chip periphery as shown in Fig.~\ref{fig:pix_ana}a.
To reduce power consumption and minimize crosstalk, the output signal amplitude is controlled by a voltage called \(\mathrm{V_{minus}}\) in Fig.~\ref{fig:pix_ana}b.
Each pixel has one corresponding hit-buffer cell that stores the timestamps of the leading and trailing edges of the comparator output in DRAM.
The time of arrival (ToA) and time over threshold (ToT) are calculated from these timestamps outside the chip.
The ToT reflects the signal amplitude, which can be estimated after applying an appropriate calibration.
The hit buffers use a column‑drain architecture.
Each column includes an end‑of‑column circuit that mediates communication between the buffer cells in that column and the finite state machine (FSM) in the chip periphery.

The chip periphery is the third main component of the chip.
It contains peripheral function blocks such as the FSM, a timestamp counter, and a newly implemented serializer.
The timestamp counter value is distributed to the hit buffers. 
It is a dual-edge counter, which improves the ToA resolution.  
The serializer encodes the data and reformats it into 32-bit packets \cite{Scherl2024}, and sends it out at the maximum speed of 1.28~Gbps.
It operates as expected and has been integrated into the digital periphery of MightyPix2 \cite{Striebig2026}.

\section{Measurement results}
\label{sec:meas}

\begin{figure}[h]
\centering 
\includegraphics[width=0.9\columnwidth]{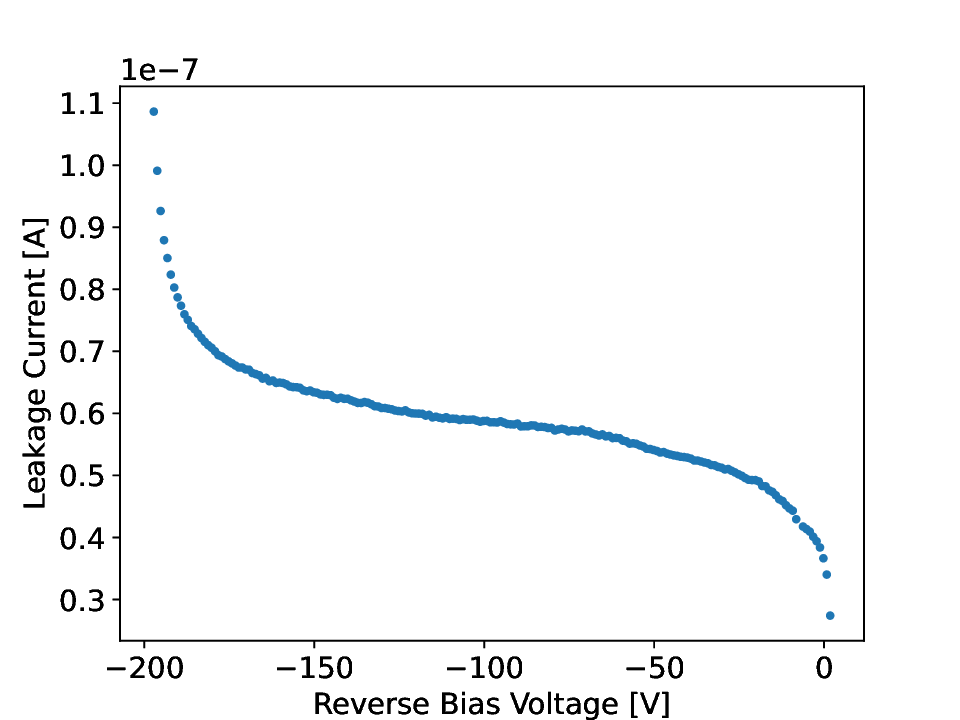}
\caption{Leakage current of LF-MightyPix as a function of the reverse bias voltage.}
\label{fig:iv}
\end{figure}

 The LF‑MightyPix has been evaluated using a test system based on the GECCO system \cite{Schimassek2021, Ehrler2021} and the Basil framework \cite{basil}.
Fig.~\ref{fig:iv} shows the leakage current as a function of the sensor bias voltage.
The leakage current begins to significantly rise at bias voltages around \SI{-195}{\volt}.
This increase suggests that the depletion region extends to the back-side of the chip, where the surface is non-uniform due to wafer thinning.
Assuming that the front‑side inactive region, i.e., from the surface to the deep n-well, and the back-side surface non-uniformity together amount to \SI{20}{\micro\meter}, the effective sensitive thickness of the sensor diode is estimated to be \SI{260}{\micro\meter}.
Using this assumption and a simple planar diode model, in which the depletion depth is  \( \mathrm{0.3 \, \sqrt{ V_{bias} \; [\si{\volt}] \;\cdot \;R_{wafer} \; [\si{\ohm\centi\meter}]}}\)\;[\si{\micro\meter}] \cite{Wermes2020}, the wafer resistivity is estimated to be \SI{4}{\kilo\ohm\centi\meter}, which is higher than the resistivity guaranteed by the foundry and is consistent with a previous study \cite{hirono2016}.
Since the required chip thickness for MightyPix is \SI{<200}{\micro\meter} (Tab. \ref{tab:req}), the results demonstrate that full depletion of the sensor is achievable below \SI{-195}{\volt}.

\begin{figure}[h] 
\centering
\includegraphics[width=0.9\columnwidth]{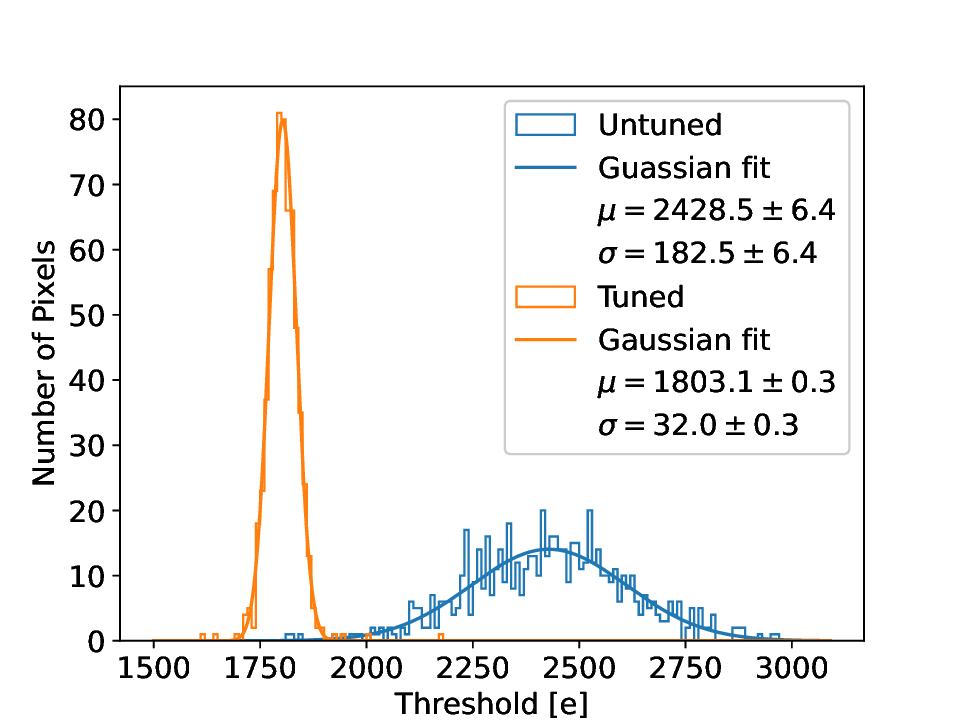}
\caption{Threshold distributions of LF-MightyPix before and after TDAC tuning. 
Each distribution is fitted with a Gaussian function, and the mean and standard deviation are indicated in the legend as \(\mu\) and \(\sigma\), respectively. The uncertainties correspond to the fitting errors.} 
\label{fig:th}
\end{figure}

TDAC tuning is performed by injecting test pulses pixel-by-pixel via \(\mathrm{C_{inj}}\), while lowering the comparator threshold in each pixel using the TDAC.
The tuned value is taken as the TDAC value at which the pixel first detects the test pulse during the threshold scan.
Fig.~\ref{fig:th} shows the intrinsic threshold distribution and the distribution after applying the TDAC tuning.
After tuning, the threshold uniformity reaches a standard deviation of \SI{32}{e^{-}}, and the mean threshold is approximately \SI{1.8}{k e^{-}}.

\begin{figure}[h]
\centering 
\includegraphics[width=0.9\columnwidth]{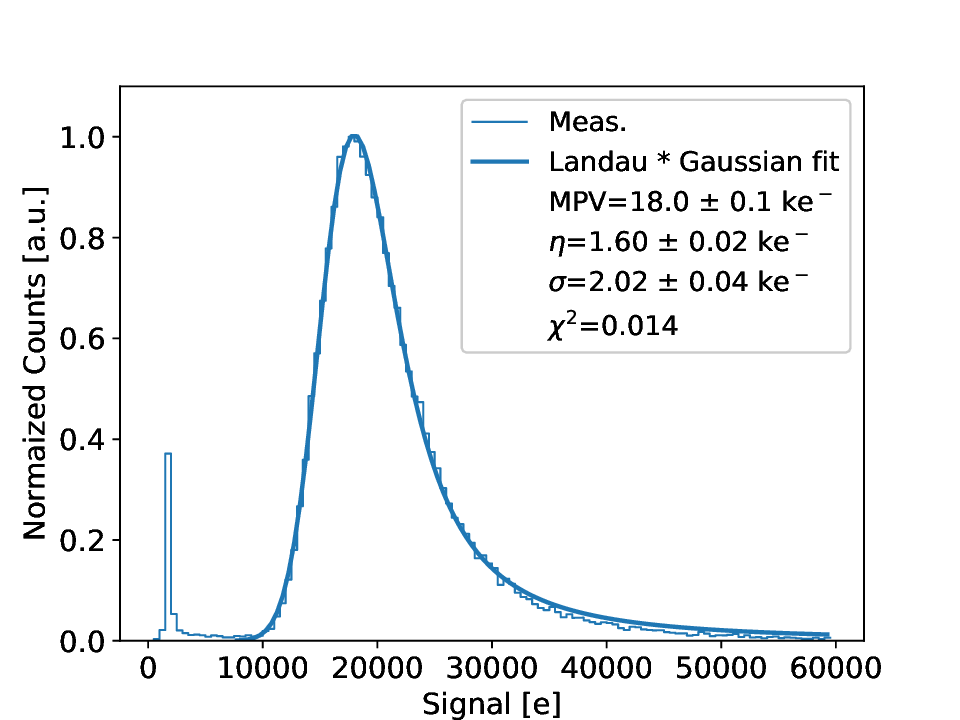}
\caption{\(\mathrm{^{90}Sr} \) spectrum measured at a bias voltage of \SI{-150}{\volt}.
The spectrum was fitted with an approximate Landau function convoluted with a Gaussian function, 
and fitting parameters are shown in the legend. \(\eta\) and \(\sigma\) are the Landau width parameter and the Gaussian standard deviation, respectively.
}
\label{fig:sr}
\end{figure}

Fig.~\ref{fig:sr} shows the \(\mathrm{^{90}Sr} \) spectrum at a bias voltage of \SI{-150}{\volt}, measured with all pixels enabled and using  the tuned TDAC values.
To mitigate the low-energy component of \(\mathrm{^{90}Sr}\), a polyethylene terephthalate filter has been inserted between the \(\mathrm{^{90}Sr} \) source and the LF-MightyPix chip, and a scintillator has been placed behind the chip as a timing reference. 
Particles occasionally produce hits that spread over multiple pixels, which typically indicates that the particle entered the sensor at a non‑perpendicular angle.
To remove such cases, the hit data are clustered by grouping hits that are contiguous in rows or columns, and only events with a cluster size of one, i.e., single‑pixel hits, are selected.
The ToT values of these single-pixel hits are then converted into signal charge.
To perform this conversion, a calibration curve is obtained for each pixel individually by injecting test pulses with varying amplitudes into \(\mathrm{C_{inj}}\) and recording the corresponding ToT response.
The measured ToT–charge relation for each pixel is subsequently used to convert the ToT values into charge in a pixel-by-pixel manner.

The most probable value (MPV) is obtained as \(18.0 \pm 0.1~\si{\kilo e^{-}}\) by fitting the spectrum with a Landau function convoluted with a Gaussian (Fig. \ref{fig:sr}). 
The small structure near \SI{0}{e^{-}} in the figure is excluded from the fit because it is an artifact caused by baseline and cross-talk noise. 
At a bias voltage of \SI{-150}{\volt}, using a wafer resistivity of \SI{4}{\kilo\ohm\centi\meter} from the leakage current measurement, the approximate depletion depth is \SI{230}{\micro\meter}. 
Assuming that \(\beta\)-particles emitted from \(\mathrm{^{90}Sr}\) are minimum ionizing particles, producing \SI{80}{e^{-}\per\micro\meter} in silicon \cite{Wermes2020}, the expected MPV of the collected charge is about \SI{18.4}{\kilo e^{-}}.
Considering the approximations and assumptions, the expected and measured MPVs show good agreement.

To evaluate the timing performance of LF‑MightyPix, the chip has been configured to the designed power consumption of \SI{11}{\micro\watt\per pixel} (\SI{110}{\milli\watt\per\centi\meter\squared}), as the time resolution depends on the power consumption.
This value is chosen to leave margin for the power consumption in the hit buffers and chip periphery to meet the requirement (Tab.~\ref{tab:req}).
The timestamp clock of LF‑MightyPix is operated at 80~MHz.  
As the leading-edge timestamp is generated by a dual-edge counter, the least significant bit (LSB) corresponds to \SI{6.25}{\nano\second}.
The measurement was performed using the same setup as for the \(\mathrm{^{90}Sr}\) spectrum.
The scintillator signal is measured using an external comparator and a time-to-digital converter operating at a sampling rate of \SI{1.28}{\giga\hertz}.

\begin{figure}[h]
\centering
\includegraphics[width=0.9\columnwidth]{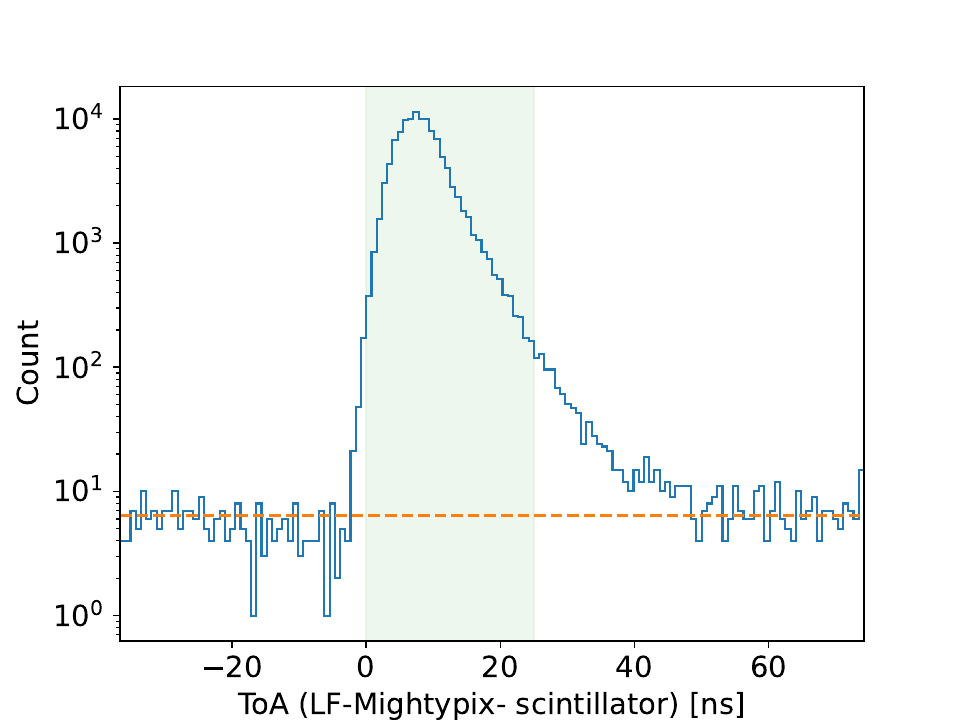}
\caption{ToA difference between LF‑MightyPix and the scintillator measured using a \(\mathrm{^{90}Sr} \) source. 
The light green hatched region indicates the \SI{25}{\nano\second} bin used to determine the in‑time percentage, defined as the fraction of events falling within this bin.
The averaged background level is shown as an orange dashed line.} 
\label{fig:toa}
\end{figure}

Fig.~\ref{fig:toa} shows the ToA difference for LF‑MightyPix at a bias voltage of \SI{-150}{\volt}, measured relative to the scintillator signal, after subtracting a fixed offset.
The hit data are clustered, and the ToA of the seed pixel, defined as the pixel with the largest ToT value in each cluster, is used as the ToA of LF-MightyPix.
A uniform background is observed in Fig.~\ref{fig:toa}.
The average value in the region where the ToA difference is less than \SI{-25}{\nano\second} is subtracted as background in the analysis, as it originates from mismatches between scintillator hits and LF‑MightyPix hits.
The scintillator is larger than the LF‑MightyPix and has no spatial resolution, while its comparator threshold is strictly set to enhance timing performance.
As a result, some scintillator events do not have corresponding LF‑MightyPix hits, and some LF‑MightyPix hits do not have corresponding scintillator events.

The ToA distribution shows different slopes on its left and right sides, with the right side falling more slowly. 
This asymmetry originates from time walk of the in‑pixel readout.
Because delay of the small signals are larger than that of the large signals, ToT‑based corrections are expected to improve the timing performance.
However, even without the ToT timing correction, the percentage of hits within a \SI{25}{\nano\second} bin has been measured as high as 99.1 \(\pm\) \SI{0.2}{\percent}.
This result confirms that LF‑MightyPix can achieve a good time resolution at the designed power consumption.

\section{Summary}
\label{sec:sum}

In this work, we present the development and evaluation of the second prototype chip of MightyPix, LF‑MightyPix, designed using the LFoundry 150 nm CMOS process for the Mighty-Tracker after the LHCb Upgrade II.
A sufficient depletion depth of the sensor (\SI{> 200}{\micro\meter}) has been confirmed by IV measurements and the \(\mathrm{^{90}Sr} \) spectrum.
A pixel-by-pixel threshold standard deviation of \SI{32}{e^-} is achieved after tuning, and 99.1~\(\pm\)~\SI{0.2}{\percent} of hits are in-time of 25~ns bin at the pixel power consumption of \SI{11}{\micro\watt\per pixel} (\SI{110}{\milli\watt\per\centi\meter\squared}).
These results show that this alternative process is a feasible option for MightyPix and also future HV-MAPS developments. 
Detection efficiency studies after irradiation are planned as a further study, as well as the development of the next prototype using this CMOS process with additional features for MightyPix.

\section*{Acknowledgments}
This work was funded by the Federal Ministry of Research, Technology and Space (BMFTR) under grant number 05H24VK4.

During the preparation of this work, the authors used Microsoft 365 Copilot to improve grammar and readability. 
After using this tool/service, the authors reviewed and edited the content as needed and take full responsibility for the content of the published article.

\bibliographystyle{elsarticle-num}
\bibliography{hdst14Rev2}

@article{HVMAPS-Peric,
      author         = "Ivan Peric",
      title          = "A novel monolithic pixelated particle detector implemented in high-voltage {CMOS} technology",

      journal       = "Nucl. Instr. and Meth. A",
      volume      = "582",
      pages        = "876-885",
      year           = "2007",
      doi         = {10.1016/j.nima.2007.07.115},
}

@book{Wermes2020,
    author = {Kolanoski, Hermann and Wermes, Norbert},
    title = {Particle Detectors: Fundamentals and Applications},
    publisher = {Oxford University Press},
    year = {2020},
    month = {06},
    isbn = {9780198858362},
    doi = {10.1093/oso/9780198858362.001.0001}
}

@techreport{LHCbScoping,
      author        = {{The LHCb collaboration}},
      title         = "{LHCb Upgrade II Scoping Document}",
      institution   = "CERN",
      reportNumber  = "CERN-LHCC-2024-010, LHCB-TDR-026",
      address       = "Geneva",
      year          = "2024"
}

@techreport{LHCb-TDR-023,
      author        = {{The LHCb Collaboration}},
      collaboration = "{LHCb} collaboration",
      title         = "{{LHCb} Framework {TDR} for the {LHCb} {Upgrade II} Opportunities in flavour physics, and beyond, in the {HL-LHC} era}",
      institution   = "CERN",
      address       = "Geneva",
      number        = "{CERN-LHCC-2021-012}",
      year          = "2022",
}

@ARTICLE{Peric2021,
  author={Peric, Ivan and et al.},
  journal={IEEE Journal of Solid-State Circuits}, 
  title={{High-Voltage} {CMOS} Active Pixel Sensor}, 
  year={2021},
  volume={56},
  number={8},
  pages={2488-2502},
  doi={10.1109/JSSC.2021.3061760}
}

@article{Hirono2020,
doi = {10.1088/1748-0221/15/05/P05013},
year = {2020},
month = {may},
publisher = {},
volume = {15},
number = {05},
pages = {P05013},
author = {Barbero, M. and et al.},
title = {Radiation hard {DMAPS} pixel sensors in 150 nm {CMOS} technology for operation at {LHC}},
journal = {Journal of Instrumentation},
}

@phdthesis{Ehrler2021,
    author       = {Ehrler, Felix Michael},
    year         = {2021},
    title        = {Characterization of monolithic {HV-CMOS} pixel sensors for particle physics experiments},
    doi          = {10.5445/IR/1000133748},
    publisher    = {{Karlsruher Institut f\"{u}}r Technologie}}

@phdthesis{Schimassek2021,
    author       = {Schimassek, Rudolf},
    year         = {2021},
    title        = {Development and Characterisation of Integrated Sensors for Particle Physics},
    doi          = {10.5445/IR/1000141412},
    publisher    = {{Karlsruher Institut f{\"{u}}r Technologie}},
    school       = {{Karlsruher Institut f{\"{u}}r Technologie}},
    language     = {english}
}

@article{Scherl2024,
doi = {10.1088/1748-0221/19/04/C04045},
year = {2024},
month = {apr},
publisher = {IOP Publishing},
volume = {19},
number = {04},
pages = {C04045},
author = {Scherl, S. and et al.},
title = {{MightyPix} at the {LHCb} {Mighty Tracker} — verification of an {HV-CMOS} pixel chip's digital readout},
journal = {Journal of Instrumentation}
}

@misc{ams, 
      author        = "{ams-OSRAM AG}",
      title         = "{ams OSRAM}",
      url           = "https://ams-osram.com/"
}

@misc{lfoundry, 
      author        = "{LFoundry S.r.l}",
      title         = "{LFoundry}",
      url           = "https://ams-osram.com/"
}

@misc{tsi, 
      author        = "{TSI Incorporated}",
      title         = "{TSI}",
      url           = "https://tsi.com/"
}

@phdthesis{Scherl2025,
    author = "Scherl, Sigrid",
    title = {Development of a Silicon Detector in {HV-CMOS} Processes for the {LHCb} {Mighty Tracker}},
    school = "University of Liverpool",
    year = "2025",
    doi = {10.17638/03190258}
}

@misc{basil, 
      author        = "{SiLab, University of Bonn}",
      title         = "Basil",
      url           = "https://github.com/silab-Bonn/basil"
}

@article{The-LHCb-Collaboration-2008,
doi = {10.1088/1748-0221/3/08/S08005},
year = {2008},
month = {aug},
publisher = {},
volume = {3},
number = {08},
pages = {S08005},
title = {The {LHCb} Detector at the {LHC}},
author = {{The {LHCb} Collaboration}},
journal = {Journal of Instrumentation}
}

@article{Striebig2026,
title = {{MightyPix} – A Novel High Voltage Monolithic Active Pixel Sensor for the proposed {LHCb} {Mighty-Tracker} },
year = {submitted},
author = {Nicolas Striebig and et al.},
journal = {Journal of Instrumentation}
}

@article{Benoit2018,
doi = {10.1088/1748-0221/13/02/P02011},
year = {2018},
month = {feb},
publisher = {},
volume = {13},
number = {02},
pages = {P02011},
author = {Benoit, M. and et al.},
title = {Testbeam results of irradiated ams {H18} {HV-CMOS} pixel sensor prototypes},
journal = {Journal of Instrumentation}
}

@article{hirono2016,
title = {{CMOS} pixel sensors on high resistive substrate for high-rate, high-radiation environments},
journal = {Nucl. Instr. and Meth. A},
volume = {831},
pages = {94-98},
year = {2016},
issn = {0168-9002},
doi = {10.1016/j.nima.2016.01.088},
author = {Toko Hirono and et al.},
}

\end{document}